\documentclass[9pt,conference]{IEEEtran}

\usepackage{waspaa25}
\usepackage{siunitx}
\usepackage{etoolbox}
\robustify\itshape
\usepackage{placeins}
\usepackage{hyperref}
\usepackage{bm} 
\usepackage{comment}

\definecolor{darkblue}{rgb}{0.0,0.0,0.6}
\definecolor{darkgreen}{rgb}{0.0,0.6,0.0}
\definecolor{pink}{rgb}{1.0, 0.2, 0.6}

\newcommand{\MG}{\textbf{MG\textsubscript{ALN}}}
\newcommand{\FM}{\textbf{FM\textsubscript{IC}}}
\newcommand{\ARXA}{\textbf{AR\textsubscript{XA}}}
\newcommand{\ARCG}{\textbf{AR\textsubscript{CG}}}

\title{Room Impulse Response Generation Conditioned on Acoustic Parameters}

\name{Silvia Arellano,$^{2}$\sthanks{Work done as part of an internship in Dolby Laboratories.}
      Chunghsin Yeh,$^{1}$
      Gautam Bhattacharya,$^{1}$
      Daniel Arteaga$^{1}$}
      
\address{$^1$ Dolby Laboratories\\
         $^2$ KTH Royal Institute of Technology
         }

\begin{document}

\maketitle

\begin{abstract}

The generation of room impulse responses (RIRs) using deep neural networks has attracted growing research interest due to its applications in virtual and augmented reality, audio postproduction, and related fields. Most existing approaches condition generative models on physical descriptions of a room, such as its size, shape, and surface materials. However, this reliance on geometric information limits their usability in scenarios where the room layout is unknown or when perceptual realism (how a space sounds to a listener) is more important than strict physical accuracy. In this study, we propose an alternative strategy: conditioning RIR generation directly on a set of RIR acoustic parameters. These parameters include various measures of reverberation time and direct sound to reverberation ratio, both broadband and bandwise. By specifying how the space should sound instead of how it should look, our method enables more flexible and perceptually driven RIR generation.
We explore both autoregressive and non-autoregressive generative models operating in the Descript Audio Codec  domain, using either discrete token sequences or continuous embeddings. Specifically, we have selected four models to evaluate: an autoregressive transformer, the MaskGIT model, a flow matching model, and a classifier-based approach. Objective and subjective evaluations are performed to compare these methods with state-of-the-art alternatives. Results show that the proposed models match or outperform state-of-the-art alternatives, with the MaskGIT model achieving the best performance. Listening examples are available at  \url{https://silviaarellanogarcia.github.io/rir-acoustic}.

\end{abstract}

\section{Introduction}
\label{sec:intro}

Room Impulse Responses (RIRs) characterize the acoustic properties of environments by capturing sound propagation from source to receiver. Accurate RIR estimation is crucial for immersive experiences in applications like virtual reality and augmented reality. In music and cinema post-production, both realistic and artistically designed RIRs are valued for their unique reverb qualities and emotional impact. 
Obtaining high-quality RIRs is challenging due to limited access to spaces, inadequate recording tools, and the difficulty of capturing RIRs throughout a room. Additionally, many RIRs do not correspond to real physical spaces. 
Consequently, RIR generation has become an important research area.

Recent advances in deep learning and generative AI have led to significant progress in room impulse response (RIR) generators,  using diverse conditioning signals and architectures. One prominent line of research focuses on the reconstruction of RIRs using visual scene representations \cite{ratnarajah2022mesh2ir,ratnarajah2024listen2scene,majumder2022few,falconperez2024novelviewacousticparameter}, where visual data helps infer local geometry. Common models include graph neural networks and conditional GANs (cGANs). However, reliance on visuals limits applicability, and a strict focus on physical realism excludes perceptually desirable but non-physical RIRs.

Another approach conditions on acoustic parameters (e.g., reverberation time) and basic room geometry. While cGANs and diffusion models have shown promising results \cite{ratnarajah2022fast,grinstein2023roomfuser}, these models still emphasize physical realism. More recently, Dellatorre et al.~\cite{dellatorre2025diffusionRIR} proposed a diffusion-based method that interpolates between measured RIRs, allowing generation of new responses from real acoustic data. However, like prior approaches, it is constrained by the need for real-world measurements and does not target the generation of perceptually desirable, non-physical RIRs.

This study presents a novel approach to generating realistic RIRs conditioned solely on acoustic parameters, without reference to room geometry.  Conditioning inputs consist on broadband and frequency-band acoustic parameters, including reverberation time, direct sound to reverberation, and source-receiver distance. This approach enables the generation of perceptually rich RIRs that need not correspond to real spaces. To our knowledge, most prior methods generating RIRs from acoustic parameters are not based on generative models: StoRIR \cite{masztalski2020storir} generates full RIRs using statistical modeling; noise-shaping methods synthesize the reverberant tail based on decay parameters \cite{noise_shaping}. Although IR-GAN \cite{ratnarajah2020ir} uses GAN to generate RIRs from acoustic parameters, it focuses on simulating far-field speech recognition, evaluating the word error rate instead of the generated RIR quality.

To evaluate different generative models, we configure our experiments around three main components:

First, we employ the Descript Audio Codec (DAC) \cite{kumar2024high} for audio tokenization. We select DAC for this study because: (1) it provides full-band high-quality audio coding, (2) it supports both discrete tokens and continuous embeddings, and (3) it encodes audio at a significantly lower frame rate, enabling efficient generative modeling.

Second, we implement a conditioning mechanism to tailor RIRs to desired acoustic parameters, which includes an optional one-hot encoding step. 
We explore two conditioning strategies: (a) integrating acoustic parameters during training using multi-layer perceptrons (MLPs) for in-context conditioning or adaptive layer normalization (adaLN) \cite{peebles2023scalable}, and (b) applying conditioning only at inference time via classifier guidance \cite{dhariwal2021}.

Third, we configure our deep neural networks to be either autoregressive (AR) or non-autoregressive (NAR) models.

\section{Acoustic parameter conditioning}
\label{sec:conditioning}

\begin{figure*}
\centering
\includegraphics[width=0.8\textwidth]{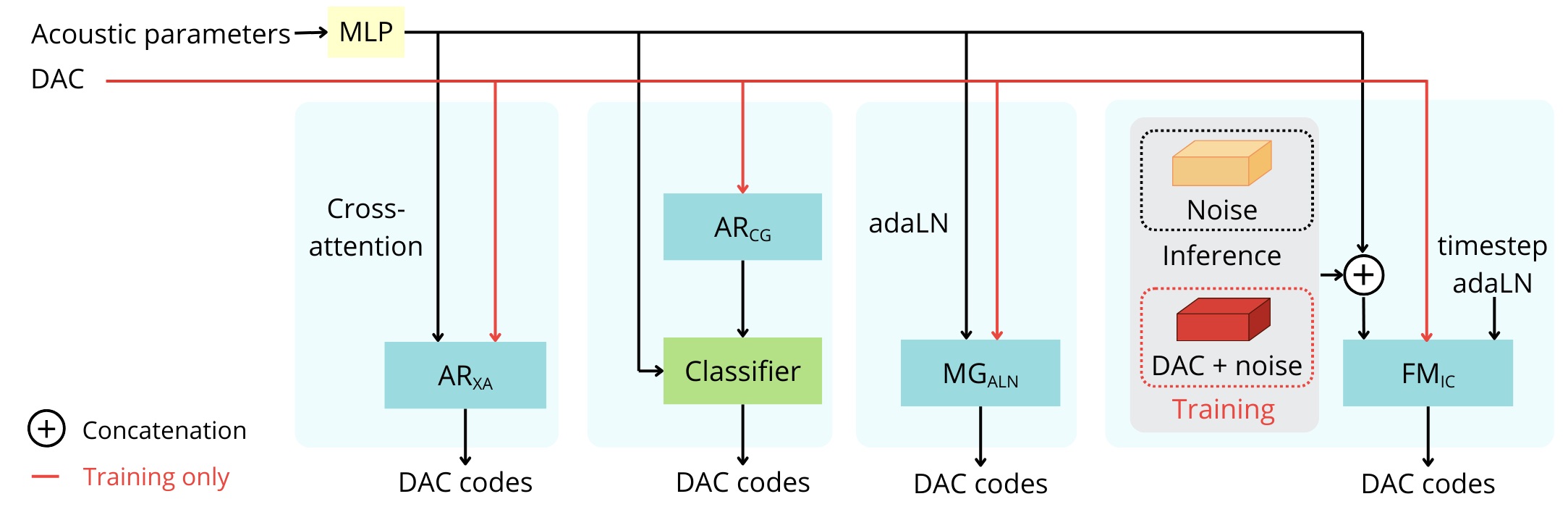}
\caption{Overview of the proposed four models with the respective conditioning mechanism. \ARXA: AR transformer with cross-attention, \ARCG: AR transformer with classifier guidance, \MG: MaskGIT with adaLN, \FM: flow matching with in-context conditioning}
\end{figure*}

The models in this work are conditioned on several acoustic parameters \cite{kuttruff_room_acoustics}: three reverberation time measures ($T_{30}$, $T_{15}$, and EDT: early decay time), two direct-sound to reverberation ratio measures ($C_{80}$ in dB and  $D_{50}$ in \%), and one distance measure (SRD: source-receiver distance). We provide both broadband and frequency-band parameters for all measures except SRD, which is specified once per sample. The frequency bands are centered at \SI{63}{\hertz}, \SI{125}{\hertz}, \SI{250}{\hertz}, \SI{500}{\hertz}, \SI{1}{\kilo\hertz}, \SI{2}{\kilo\hertz}, \SI{4}{\kilo\hertz}, and \SI{8}{\kilo\hertz}, covering the most perceptually relevant range of human hearing. In total, this results in 6 broadband parameters and $8\times5$ band-wise parameters.

We discretize the acoustic parameters into distinct classes within defined ranges. The ranges are: reverberation times ($T_{30}$, $T_{15}$, EDT) from 0.1 to \SI{1.5}{s} with 15 classes; clarity ($C_{80}$) from 0 to \SI{20}{dB} with 11 classes; definition ($D_{50}$) from 40\% to 100\% with 13 classes; and SRD from 0.3 to \SI{30}{m}  with 10 logarithmically-spaced classes. The number of classes was chosen to balance meaningful resolution with a sufficient number of data examples per class.

In addition to these core parameters, our system optionally accepts 20 parameters representing the RIR energy in mel-spaced frequency bands from \SI{20}{\hertz} to \SI{20}{\kilo\hertz}. These parameters are not used as neural network conditioning but instead serve as a spectral equalization (EQ) profile applied during postprocessing to the generated IR, ensuring that the timbre of the IR matches the desired spectral characteristics.

With the selected acoustic parameters, we study two conditioning approaches: classifier-free guidance (CFG) and classifier guidance (CG).

\subsection{Classifier-free guidance}

The acoustic parameters are treated as a 1-D conditioning tensor and passed to MLP to progressively expand the dimension to the model dimension. To enable generation control between quality and diversity, we use CFG \cite{ho2022classifier}\cite{chang2023muse} by means of introducing a learnable "unconditional" token embedding. During training, this embedding randomly replaces the conditioning embeddings with a specified probability. During inference, one can control the generation with the guidance weighting $w$: the conditional estimate weighted by $1+w$ is subtracted by the unconditional estimate weighted by $w$ to achieve the desired balance between quality and diversity.

\subsection{Classifier guidance}

We adapt classifier guidance \cite{dhariwal2021}, commonly used in diffusion models, to autoregressive models \cite{dieleman2022guidance}. Given an AR model that generates audio samples one token at a time, our goal is to produce the next code ($x_i$) such that the resulting waveform aligns with target acoustic parameters $\tau$. Using Bayes' theorem, the probability of $x_i$ is:
\begin{equation} \label{eq:classifier_guidance}
\begin{split}
\log p(x_{i}| x_{i - 1},\ldots, x_{0}; \tau_k) &= \log p(x_{i}| x_{i - 1},...x_{0}) \\
&\quad + \log p(\tau| x_{i}, x_{i - 1},\ldots, x_{0})
\end{split}
\end{equation}
The first term is the unconditional AR probability based on previous tokens, while the second term is predicted by a classifier estimating the likelihood of $\tau$ from a partial RIR ($x_0$ to $x_i$). For multiple parameter classes $\tau_k$ ($k=1,\ldots,N$), we generalize as follows:
\begin{multline} \label{eq:classifier_guidance_modif}
\log p(x_{i}| x_{i - 1},\ldots, x_{0}; \tau_1, \ldots, \tau_N) \\
\approx \lambda \log p(x_{i}| x_{i - 1},...x_{0}) + \sum^N_{k=1} w_k \log p(\tau_k| x_{i}, x_{i - 1},\ldots, x_{0})
\end{multline}
Here, $\lambda$ adjusts the influence of the AR model, and weights $w_k$ account for interdependencies among parameters ($w_k=1$ if independent; for correlated parameters, a common choice is $w_k=1/N$ \cite{Satopaa2014combining}). The classifier is queried at each generation step across all 1024 possible $x_i$ values to estimate the next-token probability conditioned on $\{\tau_k\}$.

\section{Proposed models}
\label{sec:models}

Our proposed models are commonly based on the transformer  architecture \cite{vaswani2017}, either AR or NAR. All the proposed models operate in the DAC domain, either using discrete codes or continuous latent. The DAC codes are represented as a 2-D DAC codegram $\in \mathbb{R}^{L\times T}$ where $L$ is the number of RVQ (Residual Vector Quantized) codebooks and $T$ is the number of temporal frames encoded. The AR model uses transformer decoder to generate discrete codes, whereas the NAR models use the transformer encoder to generate continuous embeddings.

\subsection{Autoregressive models}

Our AR model is a decoder-only transformer which learns to generate \textit{flattened} DAC codes. Since the DAC codes are represented like a 2-D codegram, we flatten the code into a 1-D sequence for the transformer to learn and generate like text tokens. We further propose two generation approaches based on the same AR transformer model: \ARXA\ uses cross attention for conditional generation with CFG, and \ARCG\ uses unconditional generation with CG. 

For \ARCG, a classifier is trained to predict acoustic parameter classes from partial time-domain RIRs. The architecture combines convolutional and dense layers with multiple shallow heads for parameter classification, using integrated band-pass filters for band-wise predictions.

\subsection{Non-autoregressive models}

In general, realistic RIRs are of limited duration, e.g. 2 seconds, which results in a short DAC code sequence length, allowing for practical usage of NAR models. Therefore, we propose to study two NAR models: \MG\ applies MaskGIT to discrete audio token generation, and \FM\ uses flow matching for continuous latent generation. In order to provide both options for in-context and adaLN conditioning, we adapt the transformer encoder layer as the DiT blocks described in \cite{peebles2023scalable} for the core components of the models.

In the case where DAC codes are the target, we follow the MaskGIT approach \cite{chang2022MaskGIT} to train \MG\ and generate masked audio tokens. 
Following the DAC embedding processing as described in \cite{pascual2024maskvat} we use a learnable embedding to project the discrete codes to continuous embeddings, and then sum the $L$ projected embeddings as input to the model core component. Then, the conditioning is modulated into the transformer layers via adaLN. The training/generation strategy follows that of MaskGIT with CFG \cite{pascual2024maskvat}.

In the case where DAC latent is the target, we follow the flow matching approach \cite{liu2022rectifiedflow} \cite{guan2024lafma} to train and generate continuous latent.  Unlike the above-mentioned \MG, \FM\ uses in-context conditioning (concatenate input with conditioning embeddings) for the acoustic parameters, and uses adaLN conditioning for the timesteps (sinusoidal positioning embedded). Similar to the latent flow matching approach \cite{guan2024lafma}, we train the model with MSE loss of the predicted velocity of DAC latent, and we generate the latent with Euler sampler \cite{liu2022rectifiedflow}. CFG is also applied for training/generation \cite{zheng2023guidedflows}.  In order to align the information between the discrete and the continuous representations, we use the continuous latent reconstructed from the DAC discrete codes during training.

\section{Experiments and evaluation}
\label{sec:experiments}

We present the evaluation setup, including dataset, hyperparameters, and metrics, and discuss both objective and subjective results. For comparison, we include two state-of-the-art baselines: 1) StoRIR \cite{masztalski2020storir}, which also uses acoustic parameters for conditioning and is directly comparable to our method, and 2) FastRIR \cite{ratnarajah2022fast}, a representative geometry-based approach. Both were chosen for their methodological relevance and the availability of open-source implementations.

\subsection{Dataset}

The dataset used includes both real and synthetic room impulse responses (RIRs). Real RIRs were sourced from Arni \cite{karolina2022dataset}, Motus \cite{gotz2021motus}, dEchorate \cite{dicarlo2021dechorate}, and OpenAIR \cite{murphy2010openair}, selecting only one RIR per room or furniture setup to ensure diversity. Due to the high cost and effort involved in capturing real RIRs, publicly available datasets remain limited, leading to widespread use of synthetic data. For synthetic RIR data, we used the GWA dataset \cite{tang2022gwa}. 

For each RIR in the dataset, parameters were computed from the time-domain signal, using a modified version of the Schroeder method \cite{kuttruff_room_acoustics} for the reverberation times accounting for the noise floor. Samples with high noise or manifestly invalid parameters were discarded.

The training set comprises 78,177 samples (real and synthetic), while the validation set includes 977 real RIRs. For testing, we use 50 real RIRs from the Motus dataset, chosen for their available geometric annotations.

\subsection{Implementation details}

\subsubsection*{DAC}
To ensure high quality generative modeling, we use the pretrained 44kHz DAC model, and we have also selected RIR samples with a sampling rate of at least \SI{32}{\kilo\hertz}. All the samples were truncated to \SI{2}{\second}, converted to 16-bit PCM, and resampled to \SI{44.1}{\kilo\hertz}. The audio was then encoded by DAC into 172 token frames (at \SI{86}{\hertz}) using 9 RVQ codebooks of size 1024. Tokens are used as either discrete codes or reconstructed continuous latents.

\subsubsection*{Training}
To train the models, we use adamW \cite{loshchilov2017decoupled} optimizer and cosine annealing scheduler with linear warmup\footnote{\url{https://github.com/katsura-jp/pytorch-cosine-annealing-with-warmup}}: the initial learning rate is $10^{-7}$, the warmup is $10\mathrm{k}$ steps, and the min/max learning rates are $8\times10^{-5}$ and $2\times10^{-4}$, respectively.  All the models are trained up to $200\mathrm{k}$ steps and then the best checkpoint is selected for evaluation. 

\subsubsection*{Models}
All the models use 2 transformer layers (encoder or decoder) with 16 attention heads, a model dimension of 1024, and a feedforward dimension of 4096. The conditioning network is a 3-layer MLP with interleaved fully connected layers and ReLU (rectified linear unit) activations. Notice that we use a rather shallow model in this study to match the training dataset size. 

For \ARCG\ [see \eqref{eq:classifier_guidance}], we set the hyperparameter $\lambda = 1$. Classifier weights are defined based on parameter type to account for partial correlations: $w_k = 1/\sqrt{3N_b}$ for reverberation times, $w_k = 1/\sqrt{2N_b}$ for $C_{80}$ and $D_{50}$, and $w_k = 1$ for SRD, where $N_b$ is the total number of frequency bands, including broadband. We use top-k sampling with temperature=0.5. For \ARXA, we use top-k sampling with temperature=1.

For \MG, we add a 1-D convolution layer followed by a GeLU (Gaussian error linear unit) activation to project the transformer encoder output to logits (expand from dimension $1024$ to $9\times1024$). During inference, \MG\ follows the same sampling process as described in \cite{pascual2024maskvat} with 20 sampling steps, temperature=1, diversity=0, and guidance $w$=0 for fully conditioned generation. This is meant for evaluating the effectiveness of the acoustic parameter conditioning.

For \FM, we use the one hot encoding for the acoustic parameters because our preliminary studies show that it is helpful for the in-context conditioning method used with flow matching (but not for the other models). The Euler sampler is used for inference with 25 sampling steps. We also use fully conditioned generation for evaluation.

\subsection{Evaluation metrics: objective}

\begin{table*}[bth]
\centering
\caption{Model performance compared across multiple metrics, with lower values indicating better results. Metrics include seven objective measures (in percentage points) and one perceptual measure (MUSHRA deviation from reference). Values are reported with asymmetric 95\% confidence intervals, expressed as $\text{mean}^{+\text{upper\_bound}}_{-\text{lower\_bound}}$, with bounds rounded to one significant digit and means reported to the same precision. Best-performing methods are highlighted in bold.
 } \label{tbl:summary}
\renewcommand{\arraystretch}{1.5}  

\includegraphics{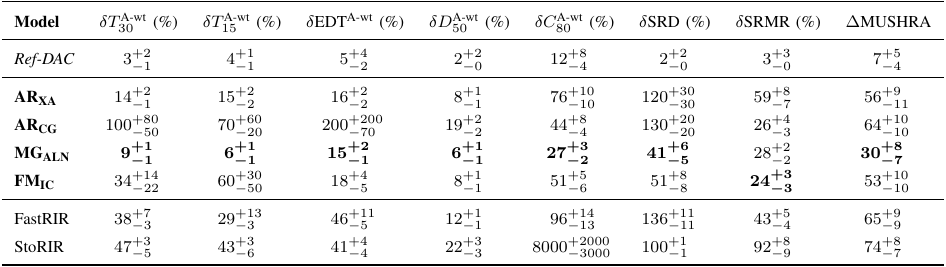}

\end{table*}

To assess each method’s ability to reconstruct acoustic parameters, we compute the relative error for each broadband and band-wise feature. For a given feature and method, the relative error $\delta\mathrm{feat}$ is defined as:
\begin{equation}\label{eq:rel_dev}
\delta\mathrm{feat} = \frac{|\mathrm{feat} - \mathrm{feat}_\text{ref}|}{\mathrm{feat}_\text{ref}}
\end{equation}
where $\mathrm{feat}$ is the predicted value and $\mathrm{feat}_\text{ref}$ is the corresponding reference value from the real RIR in the evaluation set. For $C_{80}$, which is measured in dB (a logarithmic scale), we compute the error in the linear energy domain by converting both predicted and reference values to this linear energy domain before applying the above equation.

To combine information across frequency bands, we compute a weighted average of the relative error, using A-weighting \cite{iso226_2003} to roughly reflect human hearing sensitivity. For $T_{30}$, $T_{15}$, EDT, $D_{50}$, and $C_{80}$, we compute two global metrics for each RIR: broadband and A-weighted, and report on the latter.

In addition to acoustic measures, we assess the perceptual quality of generated RIRs by convolving them with four anechoic speech samples. The Speech-to-Reverberation Modulation Energy Ratio (SRMR) \cite{falk2010srmr} is computed for both reference and generated RIRs, by comparing in each case with the anechoic samples, and the deviation  is evaluated using a relative error expression analogous to eq.\eqref{eq:rel_dev}.

To assess DAC encoding degradation, we also evaluate a version of the reference RIRs (Ref-DAC) encoded and decoded through the DAC using the same metrics.

Finally, we aggregate results by computing the mean of the evaluation metrics across samples and estimate uncertainty using bootstrapping.

\subsection{Evaluation metrics: subjective}

\begin{figure}[hbt]
    \centering
    \includegraphics[width=0.9\columnwidth]{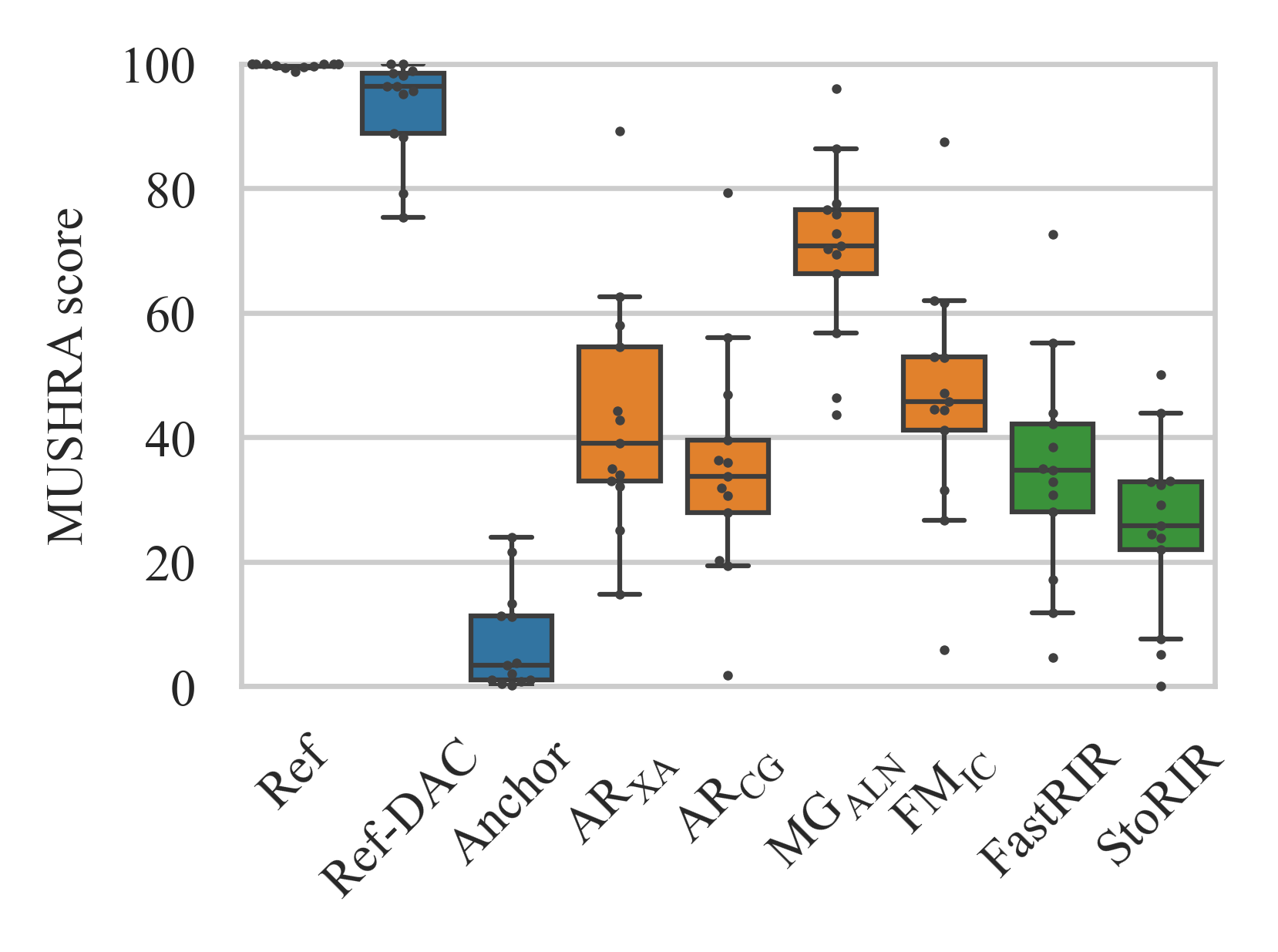}
    \caption{Boxplot of MUSHRA scores from subjective evaluations comparing proposed RIR generation methods (in orange), with references and anchor (in blue), and state-of-the-art methods (in green). Boxes show interquartile ranges, with medians marked; each dot represents a listener’s average rating. Higher scores reflect better perceptual quality.}
    \label{fig:mushra}
\end{figure}

To further assess the perceptual quality of the generated reverberation, we conducted a MUSHRA test \cite{ITU-RBS1534} with 15 participants, all with some listening experience.\footnote{Listening examples available at \url{https://silviaarellanogarcia.github.io/rir-acoustic}.} Eight real RIRs were randomly selected from the evaluation set. Each was convolved with eight dry audio signals (four transient sounds, two speech and two music samples) resulting in eight reference reverberant audio samples. The corresponding RIRs generated by each model, designed to match the acoustic features of the original, were convolved with the same dry signals to create the test reverberant audio samples. In addition to the four proposed models and two state-of-the-art baselines, the test included Ref-DAC (see above), a hidden reference, and a low-quality anchor created by convolving the dry signals with a dummy synthetic RIR consisting of \SI{500}{\milli\second} of white noise.

Participants were asked to rate how closely each test sample's reverberation matched that of the reference. Higher scores indicated closer perceptual similarity. They were instructed to focus solely on the accuracy of the reverberation match, disregarding overall audio quality or pleasantness.

\subsection{Evaluation results}

Table~\ref{tbl:summary} summarizes the evaluation results across various acoustic parameters. 
Among the proposed methods, \MG\ demonstrates the best overall performance, achieving the lowest errors in most metrics. 
\ARXA\ and \ARCG\ underperform, particularly in $T_{30}$ and $T_{15}$.
\FM\ shows mixed results, with good performance in metrics direct sound-to-reverb metrics such as $D_{50}$, but high errors in reverberation times, such as $T_{15}$. 
The state-of-the-art methods FastRIR and StoRIR demonstrate moderate and poor performance, respectively. 
Notice that most models do a relatively poor job in reproducing the SRD; however, this parameter does not have any relevance in the resulting output quality, and would be only relevant when aligning multiple RIRs.
Overall, \MG\ is the best-performing model, while StoRIR is the least effective.

The results of the MUSHRA test are presented in Fig.~\ref{fig:mushra} and also summarized in the last column of table~\ref{tbl:summary}.
Overall, the subjective evaluation aligns well with the objective test outcomes. 
Our models generally outperform the state-of-the-art methods. 
Among them, \MG\ achieves the highest MUSHRA scores, indicating the closest perceptual match to the reference. 
All pairwise comparisons are statistically significant after correction for multiple comparisons, 
except for the comparisons between \ARXA\ and FastRIR and between \ARXA\ and \FM.

The results of the evaluation also show that the degradation introduced by the DAC encoding is comparatively small, both in objective metrics and in the subjective MUSHRA scores.

\section{Conclusion and discussion}
\label{sec:conclusions}

Our experiments demonstrate that conditioning RIR generation on acoustic parameters allows our models to outperform state-of-the-art baselines, both in objective metrics and perceptual evaluations. While it is expected that we improve upon StoRIR (given its focus on data augmentation rather than perceptual quality) we find it particularly noteworthy that our models outperform FastRIR, which leverages geometric information. Importantly, each method was evaluated with its required conditioning inputs, and the same equalization procedure was applied across models to ensure a fair comparison.

Among all the evaluated models, the non-autoregressive MaskGIT-based model (\MG) stands out, achieving the best overall performance. It reached an average MUSHRA score of 70, corresponding to ``good'' quality on the MUSHRA scale.

Somewhat unexpectedly, the autoregressive models performed the weakest. In particular, \ARCG, which applies conditioning only during generation, struggled to reproduce key acoustic parameters accurately. Nonetheless, it represents a novel approach in our study. Further investigation is needed to determine whether its lower performance stems from the conditioning strategy itself, the structure of the flattened DAC codes, or the high proportion of near-silent segments in RIRs.

While the current results are promising, achieving production-level quality will require further refinements. A more diverse and higher-quality training dataset, in particular, may be key to improving perceptual performance. Additionally, we envision simplified interfaces where users do not specify all parameters directly; instead, these could be extracted from reference reverbs or controlled via intuitive mappings. Overall, this work highlights the effectiveness of conditioning on acoustic descriptors rather than geometry, demonstrates the feasibility of surpassing traditional methods, and opens new avenues for exploring conditioning strategies in RIR generation.

\section*{Acknowledgments}

We thank Giulio Cengarle for his guidance with acoustic parameters and the anonymous reviewers for their valuable feedback.

\FloatBarrier

\bibliographystyle{IEEEtran}
\bibliography{refs25}

\end{document}